\documentclass[12pt]{article}
\usepackage{latexsym}

\let\ssection=\section
\renewcommand{\section}{\setcounter{equation}{0}\ssection}

\newcommand\mathC{\mkern1mu\raise2.2pt\hbox{$\scriptscriptstyle|$}
        {\mkern-7mu\rm C}} 
\newcommand{\mathR}{{\rm I\! R}}         

\newcommand\bi{\begin{itemize}}
\newcommand\ei{\end{itemize}}
\newcommand\be{\begin{equation}}
\newcommand\ee{\end{equation}}

\begin{document}
\begin{titlepage}

\begin{center}
{\large\bf Against {\em Pointillisme} about Geometry}
\end{center}

\vspace{0.2 truecm}

\begin{center}
        J.~Butterfield\footnote{email: jb56@cus.cam.ac.uk;
            jeremy.butterfield@all-souls.oxford.ac.uk}\\[10pt] All Souls College\\ 
Oxford OX1 4AL
\end{center}

\begin{center}
       5 December 2005: a shorter version is forthcoming in {\em Proceedings of 
the 28th Ludwig Wittgenstein Symposium}, ed. F. Stadler and M. St\"{o}ltzner
\end{center}

\vspace{0.2 truecm}

\begin{abstract}
This paper forms part of a wider campaign: to deny {\em pointillisme}. That is the 
doctrine that a physical theory's fundamental quantities are defined at points of 
space or of spacetime, and represent intrinsic properties of such points or 
point-sized objects located there; so that properties of spatial or spatiotemporal 
regions and their material contents are determined by the point-by-point facts.

More specifically, this paper argues against {\em pointillisme} about the 
structure of space and-or spacetime itself, especially a paper by Bricker (1993). 
A companion paper argues against {\em pointillisme} in mechanics, especially about  
velocity; it focusses on Tooley, Robinson and Lewis.

To avoid technicalities, I conduct the argument almost entirely in the context of  
``Newtonian'' ideas about space and time. But both the debate and my arguments 
carry over to relativistic, and even quantum, physics.

\end{abstract}

\end{titlepage}

\tableofcontents

\section{Introduction}\label{Intr}
This paper forms part of a wider campaign: to deny {\em pointillisme}. That is the 
doctrine that a physical theory's fundamental quantities are defined at points of 
space or of spacetime, and represent intrinsic properties of such points or 
point-sized objects located there; so that properties of spatial or spatiotemporal 
regions and their material contents are determined by the point-by-point 
facts.\footnote{I think David Lewis first used the art-movement's name as a vivid 
label for this sort of doctrine: a precise version of which he endorsed.}

I will first describe this wider campaign (Section \ref{widercpgn}). Then I will  
argue against {\em pointillisme} as regards the structure of space and-or 
spacetime itself (Sections \ref{sec;pismintr} and \ref{ssec;allowvecs}). A 
companion paper (2006) argues against {\em pointillisme} in mechanics, especially 
as regards velocity.

I will argue that the geometrical structure of space, and-or the 
chrono-geometrical structure of spacetime, involves extrinsic properties of 
points, typically properties that I shall call `spatially extrinsic'. The main 
debate here is whether properties of a point that are represented by vectors, 
tensors, connections etc. can be intrinsic to the point; typically, {\em 
pointillistes}  argue that they can be. After formulating this debate in Section 
\ref{sec;pismintr},   I will in Section \ref{ssec;allowvecs}  focus on Bricker's 
(1993) discussion. For it is an unusually thorough {\em pointilliste} attempt to 
relate vectors and tensors  in modern geometry to the  metaphysics of properties. 
But Bricker exemplifies a tendency I reject:  the tendency  to reconcile {\em 
pointillisme} with the fact that vectorial etc. properties seem extrinsic to 
points and point-sized objects, by proposing some heterodox construal of the 
properties in question. Thus Bricker proposes that we should re-found geometry in 
terms of Abraham Robinson's non-standard analysis, which rehabilitates the 
traditional idea of infinitesimals  (Robinson 1996). I reply that once the spell 
of {\em pointillisme} is broken, such proposed heterodox foundations of geometry 
are unmotivated. 

In saying this, I do not mean to be dogmatic. I of course agree that the nature of 
the continuum is an active research area, not only historically (Mancosu (1996, 
Chapters 4f.), Leibniz 2001, Arthur 2006) but also in mathematics and philosophy. 
Indeed there are several heterodox mathematical theories of the continuum  that 
are technically impressive and philosophically suggestive. In this paper we will 
only make contact with one of them, viz.  non-standard analysis, as invoked by 
Bricker. This is for the simple reason that all the other theories offer no 
support for my target, {\em pointillisme}. More precisely: so far as I know,  
these theories  do not suggest that  fundamental quantities represent intrinsic 
properties of points or point-sized bits of matter; because either they  do not 
attribute such quantities to points, or they even deny that there are any 
points.\footnote{Broadly speaking, the second option seems more radical and worse 
for {\em pointillisme}; though in such theories, the structure of a set of points 
is often recovered by a construction, e.g. on a richly structured set of regions.}

\indent  But it is worth glimpsing at the outset the philosophical interest of 
these theories; so I here list the main ideas of some of them.\\
\indent (1): Two theories that are essentially revisions of analysis (calculus) 
are non-standard analysis, and a different rehabilitation of infinitesimals 
(smooth infinitesimal analysis; McClarty 1988, Bell 1998).\\
\indent (2): Two other approaches are based on the idea of a space with no points, 
and so are no friends of {\em pointillisme}. That is: the collection of the 
space's parts, ordered by parthood, has no atoms, i.e. no elements that themselves 
have no parts.\\
\indent \indent (i): The first is essentially a revision of measure theory, and is 
mainly motivated by its avoidance of the measure-theoretic paradoxes, like the 
Banach-Tarski paradox. (It was pioneered by Caratheodory (1963); for philosophical 
introductions, cf. Skyrms (1993), Arntzenius (2000, Section 5, pp. 201-205; 2004, 
Section 11); we will touch on the measure-theoretic paradoxes in Section 
\ref{sssec;line}, but for a full account cf. Wagon (1985).) \\
\indent \indent (ii): The second is essentially a revision of topology: topology 
is characterized by relations between regions taken as primitive. (Cf. Menger 
(1978), Roeper (1997); for a philosophical introduction, cf. Arntzenius (2004, 
Section 8-10).)

\indent Finally, three comments about the connections  between, and signficance 
of, such theories. \\
\indent  (a): These theories have various connections, which this quick list does 
not bring out. For example, Nelson (1987) shows that a modicum of non-standard 
analysis greatly simplifies a rigorous development of the theories of measure and 
probability.\\
\indent  (b): All the theories mentioned have been developed largely without 
regard to applications in physics. But Arntzenius discusses the prospects for 
doing physics, even quantum physics, in these spaces (ibid.; and for (2i), his 
2003).  (Of course, within quantum physics there is a tradition of speculation 
about discrete space or time (Kragh and Carazza 1994): for a rigorous 
non-relativistic quantum theory on a discrete space that is empirically equivalent 
to the conventional theory, cf. Davies (2003).)\\
\indent (c): As regards philosophy rather than physics, the main topic connected 
to the above theories is mereology: which has been discussed especially in 
connection with the measure-theoretic paradoxes, and (2i). Recent work includes 
Arntzenius and Hawthorne (2006, especially Sections II, IV) and  Forrest (2004, 
especially Sections 3-6; 2002, especially Sections 5-10). 

So I am very open to suggestions about heterodox treatments of the continuum. 
It is just that I find the philosophical doctrine of {\em pointillisme} an 
insufficient reason for rejecting the orthodox treatment. Similarly in my 
companion paper (2006) about mechanics; though with the difference that the 
proposals by the targeted authors, Tooley, Robinson and Lewis, do not invoke any 
well-established mathematical theory. That is: I again find {\em pointillisme} an 
insufficient reason for rejecting orthodoxy.

I will conduct the discussion almost entirely in the context of  ``Newtonian'' 
ideas about  space and time. This restriction keeps things simple: and at no cost,  
since both the debate and my arguments carry over to the treatment of space and 
time in relativistic, and even quantum, physics.

\section{The wider campaign}\label{widercpgn}
As I mentioned, this paper is part of a wider campaign, which I now sketch. I 
begin with general remarks, especially about the intrinsic-extrinsic distinction 
among properties (Section \ref{ssec;connect}). Then I state my main claims; first 
in brief (Section \ref{ssec;apfp}), then in more detail (Section 
\ref{ssec;moredetail}).

\subsection{Connecting physics and metaphysics}\label{ssec;connect}
My wider campaign aims to connect what modern classical physics says about matter 
with two debates in modern analytic metaphysics. The first debate is about {\em 
pointillisme}; but understood as a metaphysical doctrine rather than a property of 
a physical theory. So, roughly speaking, it is the debate whether the world is 
fully described  by all the intrinsic properties of all the points and-or 
point-sized bits of matter. The second debate is whether an object persists over 
time by the selfsame object existing at different times (nowadays called  `{\em 
endurance}'), or by different temporal parts, or stages, existing at different 
times (called `{\em perdurance}').

Endeavouring to connect classical physics and metaphysics raises two large initial 
questions  of philosophical method. What role, if any, should the results of 
science have in metaphysics? And supposing metaphysics should in some way 
accommodate these results, the fact that we live (apparently!) in a  quantum 
universe prompts the question why we should take classical physics to have any 
bearing on metaphysics. I address these questions in my (2004: Section 2, 2006a: 
Section 2). Here I just summarize my answers.\\
\indent  I of course defend the relevance  of the results of science for 
metaphysics; at least for that branch of it, the philosophy of nature, which 
considers such notions as space, time matter and causality. And this includes 
classical physics, for two reasons.\\
\indent First, much analytic philosophy of nature assumes, or examines, so-called 
`common-sense' aspects and versions of these notions: aspects and versions which 
reflect   classical physics, especially mechanics, at least as taught in 
high-school or elementary university courses. One obvious example is modern 
metaphysicians' frequent discussions of matter as point-particles, i.e. 
extensionless point-masses moving in a void (and so interacting by 
action-at-a-distance forces), or as continua, i.e. bodies whose entire volume, 
even on the smallest scales, is filled with matter. Of course, both notions arose 
in mechanics in the seventeenth and eighteenth century.\\
\indent Second, classical physical  theories, in particular mechanics, are much 
more philosophically suggestive, indeed subtle and problematic, than philosophers  
generally realize. Again, point-particles and continua provide examples. The idea 
of  mass concentrated in a spatial point (indeed, different amounts at different 
points) is, to put it mildly, odd; as is action-at-a-distance interaction. And 
there are considerable conceptual tensions in the mechanics of continua; (Wilson 
(1998) is a philosopher's introduction). Unsurprisingly, these subtleties and 
problems were debated in the heyday of classical physics, from 1700 to 1900; and 
these debates had an enormous influence on philosophy through figures like Duhem, 
Hertz and Mach---to mention only figures around 1900 whose work directly 
influenced the analytic tradition. But after the quantum and relativity 
revolutions, foundational issues in classical mechanics were largely ignored, by 
physicists and mathematicians as well as by philosophers. Besides, the growth of 
academic philosophy after 1950 divided the discipline into  compartments, labelled 
`metaphysics', `philosophy of science' etc., with the inevitable result that there 
was less communication between, than within, compartments.\footnote{Thus I see my 
campaign as a foray into the borderlands between metaphysics and philosophy of 
physics: a territory that I like to think of as inviting exploration, since it 
promises to give new and illuminating perspectives on the theories and views of  
the two communities lying to either side of it---rather than as a no-man's-land 
well-mined by two sides, ignorant and suspicious of each other!}

Setting aside issues of philosophical method, {\em pointillisme} and persistence 
are clearly large topics; and each is the larger for being treatable using the 
very diverse methods and perspectives of both disciplines, metaphysics and 
physics. So my campaign has to be selective in the ideas I discuss and in the 
authors  I cite. Fortunately, I can avoid several philosophical controversies, and 
almost all technicalities of physics.\footnote{Though persistence is not this 
paper's topic, I note that among the philosophical issues my campaign avoids are 
several about persistence, such as: (a) the gain and loss of parts (as in Theseus' 
ship); (b) the relation of ``constitution'' between matter and object (as in the 
clay and the statue); (c) vagueness, and whether there are vague objects. Agreed, 
there are of course  connections between my claims and arguments, and the various 
issues, both philosophical and physical, that I avoid: connections which it would 
be a good project to explore. But not in one paper, or even in one campaign!}

But I need to give at the outset some details about how I avoid philosophical 
controversy about the intrinsic-extrinsic distinction among properties, and about 
how this distinction differs from three that are prominent in mathematics and 
physics.

\subsubsection{Avoiding controversy about the intrinsic-extrinsic 
distinction}\label{sssec;ied}
My campaign does not need to take sides in the ongoing controversy about how to 
analyse, indeed understand, the intrinsic-extrinsic distinction. (For an 
introduction, cf. Weatherson (2002, especially Section 3.1), and the symposium, 
e.g. Lewis (2001), that he cites.) Indeed, most of my discussion can make do with 
a much clearer distinction, between what Lewis (1983, p. 114) dubbed the `positive 
extrinsic' properties, and the rest. This goes as follows. \\
\indent Lewis was criticizing Kim's proposal, to analyze extrinsic properties as 
those that imply {\em  accompaniment}, where something is accompanied iff it 
coexists with some wholly distinct contingent object, and so to analyze intrinsic 
(i.e. not extrinsic) properties as those that are compatible with being 
unaccompanied, i.e. being the only contingent object in the universe (for short: 
being {\em lonely}). Lewis objected that loneliness is itself obviously extrinsic. 
He also argued that there was little hope of amending Kim's analysis. In 
particular, you might suggest that to be extrinsic, a property must either imply 
accompaniment or imply loneliness: so Lewis dubs these disjuncts `positive 
extrinsic' and `negative extrinsic' respectively. But Lewis points out that by 
disjoining and conjoining properties, we can find countless extrinsic properties 
that are neither positive extrinsic nor negative extrinsic; (though `almost any 
extrinsic property that a sensible person would ever mention is positive 
extrinsic' (1983, p. 115)).\\
\indent  This critique of Kim served as a springboard: both for Lewis' own 
preferred analysis, using a primitive notion of naturalness which did other 
important work in his metaphysics (Lewis 1983a); and for other, metaphysically 
less committed, analyses, developed  by Lewis and others (e.g. Langton and Lewis 
1998, Lewis 2001).\\
\indent But I will not need to pursue these details. As I said, most of my 
campaign can make do with the notion of positive extrinsicality, i.e. implying 
accompaniment, and its negation. That is, I can mostly take {\em pointillisme} to 
advocate properties that are intrinsic in the weak sense of not positively 
extrinsic. So this makes my campaign's claims, i.e. my denial of {\em 
pointillisme}, logically stronger; and so I hope more interesting. Anyway, my 
campaign makes some novel proposals about positive extrinsicality: in this paper, 
I distinguish temporal and spatial (positive) extrinsicality; and in the companion 
paper against {\em pointillisme} in mechanics, I propose degrees of (positive) 
extrinsicality.

\subsubsection{Distinction from three mathematical 
distinctions}\label{sssec;3maths}
Both the murky intrinsic-extrinsic distinction, and the clearer distinction 
between positive extrinsics and the rest, are different distinctions from three 
that are made within mathematics and physics, especially in those parts relevant 
to us: viz. pure and applied differential geometry. The first of these 
distinctions goes by the name `intrinsic'/`extrinsic'; the second is called 
`scalar'/`non-scalar', and the third is called `local'/`non-local'. They are as 
follows.

\indent (i): The use of `intrinsic' in differential geometry is a use which is 
common across all of mathematics: a feature is intrinsic to a mathematical object 
(structure) if it is determined (defined) by just the object as given, without 
appeal to anything extraneous---in particular a choice of a coordinate system, or 
of a basis of some vector space, or of an embedding of the object into another. 
For example, we thus say that the intrinsic geometry of a cylinder is flat; it is 
only as embedded in $\mathR^3$ that it is curved.

\indent (ii): Differential geometry classifies quantities according to how they 
transform between coordinate systems: the simplest case being scalars which have 
the same value in all coordinate systems. (Nevermind the details of how the other 
cases---vectors, tensors, connections, spinors etc.---transform.)

\indent (iii): Differential geometry uses `local' (as vs. `global') in various 
ways. But the central use is that a mathematical object (structure) is local if it 
is associated with a point by being determined (defined) by the mathematical  
structures defined on {\em any} neighbourhood, no matter how small, of the point. 
In this way, the instantaneous velocity of a point-particle at a spacetime point, 
and all the higher derivatives of its velocity, are local since their existence 
and values are determined by the particle's trajectory in an arbitrarily small 
neighbourhood of the point. Similarly, an equation is called `local' if it 
involves only local quantities. In particular, an equation of motion is called 
`local in time' if it describes the evolution of the state of the system at time 
$t$ without appealing to any facts that are a finite (though maybe very small) 
time-interval to the past or future of $t$.

I will not spell out {\em seriatim} some examples showing that the two 
philosophical distinctions are different from the three mathematical ones. Given 
some lessons in differential geometry (not least learning to distinguish (i) to 
(iii) themselves!), providing such examples is straightforward work. Suffice it to 
make three comments, of increasing relevance for this paper.

(1): It would be a good project to explore the detailed relations between these 
distinctions. In particular, the mathematical distinction (i) invites comparison 
with Vallentyne's (1997) proposal about the intrinsic-extrinsic distinction. 
Besides, there are yet other distinctions to explore and compare: for example, 
Earman (1987) catalogues some dozen senses of `locality'.\\
\indent But in this paper and its companion, two of the various differences 
amongst these distinctions are especially relevant.

(2): The first is the difference between mathematical locality, (iii) above, and 
philosophical intrinsicality. The difference is clear for the case of 
instantaneous velocity. This is the main topic of my (2006); but the idea is that 
velocity has implications about the object at other times, for example that it 
persists for some time. So most philosophers say that instantaneous velocity is an 
extrinsic property.  I agree. But emphasising its extrinsicness tends to make one 
ignore the fact that it is mathematically local, i.e. determined by the object's 
trajectory in an arbitrarily small time-interval. And in pure and applied 
differential geometry, it would be hard to over-estimate the importance of---and 
practitioners' preference for!---such local quantities and local equations 
involving them. (It is this locality that prompts me to speak of instantaneous 
velocity (and other local quantities) as `hardly extrinsic'.)
 
\indent (3): In this paper, we will also note the difference between being a 
mathematical scalar, (ii) above, and being philosophically intrinsic.  Thus 
philosophers tend  to think that any scalar quantity represents an intrinsic 
property of the points on which it is defined; (so that the {\em pointilliste} has 
only to worry about whether vectors, tensors etc. can represent intrinsic 
properties). But as we shall see in Section \ref{344A;vslocaldupls}, that is 
wrong. For the scalar curvature $R$ at a point $p$ is surely extrinsic in the 
philosophical  sense, since it gives information about the geometry of 
neighbourhoods of $p$. ($R$ is also local and mathematically intrinsic; i.e. on 
the ``intrinsic side'' of all three mathematical distinctions, (i)-(iii).)

\subsection{Classical mechanics is not {\em pointilliste}, and can be  
perdurantist}\label{ssec;apfp}
\subsubsection{Two versions of {\em pointillisme}}\label{sssec;2versions}
To state my campaign's main claims, it is convenient to first distinguish a weaker 
and a stronger version of {\em pointillisme}, understood as a metaphysical 
dosctrine. They differ, in effect, by taking `point' in {\em pointillisme} to 
mean, respectively, spatial,  or spacetime,  point.

 Taking `point' to mean `spatial point', I shall take {\em pointillisme} to be, 
roughly, the doctrine that the instantaneous state of the world is fully described  
by all the intrinsic properties, at that time, of all spatial points and-or 
point-sized bits of matter.\\
\indent As I said in Section \ref{ssec;connect}, my campaign can mostly take 
`intrinsic' to mean `lacking implications about some wholly distinct contingent 
object'; in other words, to mean the negation of Lewis' `positive extrinsic' (i.e. 
his `implying accompaniment'). But for this version of {\em pointillisme}, I will 
take `intrinsic' to mean `{\em spatially} intrinsic'. That is,  attributing such a 
property to an object carries no implications about spatially distant objects; but 
it {\em can} carry implications about objects at other times. (Such objects might 
be other temporal parts of the given object.) So I shall call this version, `{\em 
pointillisme} as regards space'.

On the other hand: taking `point' to mean `spacetime point', I shall take {\em 
pointillisme} to be, roughly, the doctrine that the history of the world is fully 
described  by all the intrinsic properties of all the spacetime  points and-or all 
the intrinsic properties at all the various times of point-sized bits of matter 
(either point-particles, or in a continuum). And here I take `intrinsic' to mean 
just the negation of Lewis' `positive extrinsic'. That is, it means `both 
spatially and temporally intrinsic': attributing such a property carries no 
implications about objects at other places, or at other times. I shall call this 
stronger version, `{\em pointillisme} as regards spacetime'.

So to sum up: {\em pointillisme} as regards space vetoes spatial extrinsicality; 
but {\em pointillisme} as regards spacetime also vetoes temporal extrinsicality.

On either reading of {\em pointillisme}, it is of course a delicate matter to 
relate such metaphysical doctrines, or the endurance-perdurance debate,  to the 
content of specific physical theories. Even apart from Section 
\ref{ssec;connect}'s questions  of philosophical method, one naturally asks, for 
example, how  philosophers' idea of  intrinsic property relates to the idea of a 
physical quantity. For the most part, I shall state my verdicts about  such 
questions case by case. But one main tactic for relating the metaphysics to the 
physics will be to formulate {\em pointillisme} as a doctrine relativized to (i.e. 
as a property of) a given physical theory (from Section \ref{ssec;moredetail} 
onwards). Anyway, I can already state my main claims, in terms of these two 
versions of {\em pointillisme}. More precisely, I will state them as denials of 
two claims that are, I think,  common in contemporary metaphysics of nature.

\subsubsection{Two common claims}\label{sssec;2claims}
Though I have not made a survey of analytic metaphysicians, I think many of them 
hold two theses, which I will dub (FPo) (for `For {\em Pointillisme}') and (APe) 
(for `Against perdurantism'); as follows.

\indent (FPo): Classical physics---or more specifically, classical 
mechanics---supports {\em pointillisme}: at least as regards  space, though  
perhaps not as regards spacetime. There are two points here:---\\
\indent \indent (a): Classical physics is free of various kinds of  ``holism'', 
and thereby anti-{\em pointillisme}, that are suggested by quantum theory. Or at  
least: classical  mechanics is free. (With the weaker claim, one could allow, and 
so set aside, some apparently anti-{\em pointilliste} features of advanced 
classical physics, e.g. anholonomies in electromagnetism and the 
non-localizability of gravitational energy in general relativity: features rich in 
philosophical suggestions (Batterman 2003, Belot 1998, Hoefer 2000)---but not for 
this paper!)\\
\indent \indent (b): The concession, `perhaps  not as regards spacetime', arises 
from the endurance-perdurance debate. For it seems  that {\em pointillisme} as 
regards spacetime must construe persistence as perdurance;  (while {\em 
pointillisme} as regards space could construe it as endurance). And a well-known 
argument, often called `the rotating discs argument', suggests that perdurance  
clashes with facts about the rotation of a continuum (i.e. a continuous body) in 
classical mechanics. So the argument suggests that classical mechanics must be 
understood as ``endurantist''. Besides, whether or not one endorses the argument,  
in classical mechanics the persistence  of objects surely {\em can} be understood 
as endurance---which conflicts with {\em pointillisme} as regards spacetime.\\
\indent  (The considerations under (a) and (b) are usually taken as applying 
equally well to non-relativistic and  relativistic classical mechanics: an 
assumption I largely endorse.)

I also think that many metaphysicians would go further and hold that:\\
\indent (APe): Classical mechanics does indeed exclude {\em pointillisme} as 
regards spacetime: their reason being that this {\em pointillisme} requires 
perdurance and that they endorse the rotating discs argument. So they hold that in 
classical mechanics the persistence  of objects  {\em must} be understood as 
endurance, and that this forbids {\em pointillisme} as regards spacetime. 

\subsubsection{My contrary claims}\label{sssec;my2claims}
I can now state the main position of my wider campaign. Namely, I {\em deny} both 
claims, (FPo) and (APe), of Section \ref{sssec;2claims}. I  argue for two contrary 
claims, (APo) (for `Against {\em Pointillisme}) and (FPe) (for `For 
perdurantism'), as follows.

\indent (APo): Classical mechanics does {\em not} support {\em pointillisme}.\\
\indent By this I do not mean just that:\\
\indent\indent (a) it excludes {\em pointillisme} as regards spacetime.\\
 Nor do I just mean:\\
\indent\indent (b) it allows one to construe the persistence  of objects as 
endurance.\\
 (But I agree with both (a) and (b).)  Rather, I also claim: classical mechanics  
excludes {\em pointillisme} as regards space. That is: it needs to attribute 
spatially extrinsic properties to spatial points, and-or point-sized bits of 
matter. (But this will not be analogous to the kinds of ``holism'' suggested by 
quantum theory.)
 
\indent (FPe): Though (as agreed in (APo)) classical mechanics  excludes {\em 
pointillisme} as regards spacetime (indeed, also: as regards space): classical 
mechanics is {\em compatible} with perdurance.  That is: despite the rotating 
discs argument,  one {\em can} be a ``perdurantist'' about the persistence of 
objects in classical mechanics. The reason is that once we reject {\em 
pointillisme}, perdurance does not need persistence to supervene on temporally 
intrinsic facts. In fact, perdurantism can be defended by swallowing just a small 
dose of temporal extrinsicality. 
  
So to sum up my wider campaign, I claim that:---\\
\indent (APo): Classical mechanics denies {\em pointillisme}, as regards space as 
well as spacetime. For it needs to use spatially extrinsic properties of spatial 
points and-or point-sized bits of matter, more than is commonly believed. \\
\indent (FPe): Classical mechanics permits perdurantism. It does not require 
temporally extrinsic properties (of matter, or objects), in the sense of requiring 
persistence to be endurance: as is commonly believed. A mild dose of temporal 
extrinsicality can reconcile classical mechanics with perdurance.

To put the point in the philosophy of mind's terminology of `wide' and `narrow' 
states, meaning (roughly) extrinsic and intrinsic states, respectively: I maintain 
that classical mechanics:\\
\indent (APo): needs to use states that are spatially wide, more than is commonly 
believed; and \\
\indent (FPe): does not require a specific strong form of temporal width, viz. 
endurance. With a small dose of temporal extrinsicality, it can make do with 
temporally quite narrow states---and can construe persistence as perdurance.

\subsection{In more detail ...}\label{ssec;moredetail} 
So much by way of an opening statement.  I will now spell out my main claims in a 
bit more detail: first (APo), and then, more briefly, (FPe).

\paragraph{2.3.A Four violations of {\em pointillisme}}\label{sssec;fourviol}
I will begin by stating {\em pointillisme}  as a trio of claims that apply to any 
physical theory; and making two comments. Then I list four ways in which 
(chrono)-geometry and classical  mechanics violate {\em pointillisme}: three will 
form the main topics of this paper and its companion.

 The trio of claims is as follows:\\
\indent (a): the fundamental quantities of the physical theory in question are to 
be defined at points of space or of spacetime;\\
\indent (b): these quantities represent intrinsic properties of such points; \\
\indent (c): models of the theory---i.e. in physicists' jargon, solutions of its 
equations, and in metaphysicians' jargon, possible worlds according to the 
theory---are fully defined by a specification of the quantities' values at all 
such points.\\
\indent So, putting (a)-(c) together: the idea of {\em pointillisme} is that the 
theory's  models (or solutions or worlds) are something like conjunctions or 
mereological fusions of ``ultralocal facts'', i.e. facts at points.

Two comments. First: the disjunction in (a), `at points of space or of spacetime', 
corresponds to   Section \ref{ssec;apfp}'s distinction between {\em pointillisme} 
as regards space, and as regards spacetime. Nevermind that it does not imply the 
convention I adopted in Section \ref{ssec;apfp}, that {\em pointillisme} as 
regards spacetime is a {\em stronger} doctrine: since it vetoes temporally 
extrinsic properties, {\em as well as} spatially extrinsic ones. The context will 
always make it clear whether I mean space or spacetime (or both); and whether I 
mean spatially or temporally extrinsic (or both).\\
\indent Second: Though I have not made a systematic survey, there is no doubt that 
{\em pointillisme}, especially its claims (a) and (b), is prominent in 
contemporary metaphysics of nature, especially of neo-Humean stripe. The prime 
example is David Lewis' metaphysical system, which is so impressive in its scope 
and detail.  One of his main metaphysical theses, called `Humean supervenience', 
is a version of {\em pointillisme}: I will return to it in Section 
\ref{ssec;HSintr}. 

When we apply (a)-(c) to classical mechanics, there are, I believe, four main ways 
in which {\em pointillisme} fails: or, more kindly expressed, four concessions 
which {\em pointillisme} needs to make. The first three violations (concessions) 
occur in the classical mechanics both of point-particles and of continua; the 
fourth is specific to continua. And the first two  are addressed in this paper; 
the third is discussed in the companion paper (2006).

(1): The first violation is obvious and minor. Whether matter is conceived as 
point-particles or as continua, classical  mechanics uses a binary relation of 
occupation, `... occupies ...', between bits of matter and spatial or spacetime 
points (or, for extended parts of a continuum: spatial or spacetime regions). And 
this binary relation presumably brings with it extrinsic properties of its relata: 
it seems an extrinsic property of a point-particle (or a continuum, i.e. a 
continuous body) that it occupy a certain spatial or spacetime point or region; 
and conversely.

Agreed, there is more to be said about this claim (as always in philosophy!): both 
about (a) the connections between the intrinsic-extrinsic distinction among 
properties and the classification of relations, and (b) how the individuation of 
spatial or spacetime points or regions might depend on matter (the ``relational 
conception'' of space or spacetime). I will  discuss  (a) and (b), albeit briefly 
in Sections \ref{ssec;allowspace}. But anyway, I will  there endorse the claim. 
That is: the concession will remain in force: the {\em pointilliste} about 
classical mechanics  should accept this binary relation of occupation, and the 
modicum of extrinsicality it involves.
   
\indent (2): Classical mechanics (like other physical theories) postulates 
structure for space and-or spacetime (geometry or chrono-geometry); and this 
involves a complex network of geometric  relations between, and so extrinsic 
properties of, points. This concession is of course more striking as regards space 
than time: three-dimensional Euclidean geometry involves more structure than does 
the real line. This will be the main topic of this paper.

 (3): Mechanics needs of course to refer to the instantaneous  velocity or 
momentum of a body; and this is temporally extrinsic to the instant  in question, 
since for example it implies the body's existence at other times. (But it is also 
local in the sense of (iii), Section \ref{sssec;3maths}.)  So this second 
violation imposes temporal, rather than spatial, extrinsicality; i.e.  
implications about other times, rather than other places.\\
\indent This is the main topic of Butterfield (2006). But I should stress here 
that this third violation is {\em mitigated} for point-particles, as against 
continua. For a {\em pointilliste} {\em can} maintain that the persistence of 
point-particles supervenes on facts that, apart from the other violations (i.e. 
about `occupies' and (chrono)-geometry), are {\em pointillistically} acceptable: 
viz. temporally intrinsic facts about which spacetime  points are occupied by 
matter. In figurative terms: the void between distinct point-particles allows one 
to construe their persistence in terms of tracing the curves in spacetime 
connecting points that are occupied by matter. I develop this theme in my (2005). 
On the other hand: for a continuous body, the persistence of spatial parts 
(whether extensionless or extended) does {\em not} supervene on such temporally 
intrinsic facts. This is the core idea of  the rotating discs argument, mentioned 
in Section \ref{sssec;2claims}.\\
\indent To sum up: the rotating discs argument means that {\em pointillisme} fits 
better with point-particles than with continua. To put the issue in terms of 
Section \ref{ssec;apfp}'s two forms of {\em pointillisme}: the strong form of {\em 
pointillisme}, {\em pointillisme} as regards spacetime, fails for the classical 
mechanics of continua, even apart from the other concessions mentioned. 

(4): Finally, there is a fourth way that the classical mechanics of continua 
violates {\em pointillisme}: i.e., a fourth concession that {\em pointillisme} 
needs to make. Unlike the rotating discs argument, this violation seems never to 
have been noticed in recent analytic metaphysics; though the relevant physics goes 
back to Euler. Namely, the classical mechanics of continua violates (the weaker 
doctrine of) {\em pointillisme} as regards space, because it must be formulated in 
terms of spatially extended regions and their properties and relations. But in 
this paper, I set this fourth violation aside entirely; my (2006a) gives details.
  
So to sum up these four violations, I claim (APo): classical mechanics violates 
{\em pointillisme}. This is so even for the weaker doctrine, {\em pointillisme}  
as regards space. And it is especially so, for the classical mechanics of continua 
rather than point-particles.

\paragraph{2.3.B For perdurantism}\label{sssec;forperdm}
I turn to Section \ref{sssec;my2claims}'s second claim, (FPe):  that once {\em 
pointillisme} is rejected, perdurantism does not need persistence to supervene on 
temporally intrinsic facts, and can be defended for classical mechanics provided 
it swallows a small  dose of temporal extrinsicality. 

About (FPe) I can be much briefer, since this paper will not need details. I will 
just identify this small dose: it is the extrinsicality of the third violation of 
{\em pointillisme} above---in particular,  the presupposition of persistence by 
the notion of a body's instantaneous velocity. Thanks to the rotating discs 
argument, `body' here means especially `point-sized bit of matter in a continuum'; 
since for point-particles we can construe persistence as perdurance without having 
to take this dose. Elsewhere (2004, 2004a, 2006) I argue that for a ``naturalist'' 
perdurantist, this dose is small enough to swallow.

\section{Can properties represented by vectors be intrinsic to a 
point?}\label{sec;pismintr}
\subsection{Prospectus}\label{ssec;pismprospect}
I turn to the geometrical structure of space, and-or the chrono-geometrical 
structure of spacetime. I will argue that this  structure involves extrinsic 
properties, especially spatially extrinsic properties. I will undertake three 
specific tasks, in Sections \ref{ssec;HSintr}, \ref{ssec;allowspace}, 
\ref{ssec;allowvecs} respectively.

 In Section \ref{ssec;HSintr}, I present Lewis' version of {\em pointillisme}. 
Though this version is in some ways logically stronger than I need, it is 
important  to present  it. Not only has it been  a focus of recent metaphysical 
discussion; it is also needed for Section \ref{ssec;allowvecs}.  

In Sections \ref{ssec;allowspace} and \ref{ssec;allowvecs}, I argue in detail that 
{\em pointillisme} needs to be qualified to accommodate  the structure of space 
and-or spacetime.  I think the need for this qualification is uncontentious; in 
particular, it is agreed by Lewis. But how exactly to state the qualification is a 
matter that is both important and unresolved.\\
\indent It is important for three reasons. First, all physical theories of course  
appeal to space (and-or spacetime). Second, they  all represent  the properties 
that encode the structure of space or spacetime, with mathematical entities such 
as vectors, tensors, connections etc. So the question arises: can properties that 
are so represented be intrinsic to a point? The third reason is taken up in the 
companion paper (2006): it is that physical  theories also represent the {\em 
other} properties they mention, i.e. properties of matter such as velocity, 
momentum etc., by such mathematical entities as vectors, tensors, connections etc.

So the question---{\em can properties represented by vectors, tensors etc. be 
intrinsic to a point?}---is at the centre of this paper (and its companion). 
First, in Section \ref{ssec;allowspace}, I will lead up to this question by 
discussing, in a broadly metaphysical way, how to represent the structure of space 
or spacetime. (I will concentrate on the notion of length, and so on space rather 
than spacetime; but this discussion carries over intact to the case of spacetime.) 
Once the question is posed, Section \ref{ssec;allowvecs} addresses it in detail, 
using as a foil Bricker's  (1993).

As I said in Section \ref{Intr}, Bricker's paper illustrates how strongly some 
contemporary metaphysicians are attracted by {\em pointillisme}. For recognizing 
that they must accept vectorial properties in physical theories, and that these 
{\em seem} not to be intrinsic to points, they propose to save {\em pointillisme} 
by advocating a heterodox construal of the property. Thus in Section 
\ref{ssec;allowvecs}, Bricker will construe the metric tensor of differential 
geometry in terms of non-standard analysis. (And in the companion paper,  Tooley 
and others will construe  instantaneous velocity as intrinsic.)  My own view will 
of course be that there is no need for such heterodoxy: instead, we can and should 
reject {\em pointillisme}.

My discussion will be simplified by a restriction. I will consider only properties 
represented by vectors and tensors, which I will for short call {\em vectorial 
properties} and  {\em tensorial properties}: not those represented by other 
mathematical entities such as connections. This restriction will be natural, in 
that:\\
\indent (i) vectors and tensors are about the simplest of the various mathematical 
entities that physical theories use to represent properties and relations---so 
they are the first case to consider;\\  
\indent (ii) the restriction is common in the literature; in fact most of the 
authors I discuss (here and in the companion paper) consider only vectorial 
properties.

\subsection{Humean supervenience}\label{ssec;HSintr}
I will assume familiarity with the main ideas of Lewis' metaphysical system, above 
all his notions of possible world and natural property.
Central to this system is Lewis' version of {\em pointillisme}, which he says 
(1994, p. 494)  is inspired by classical physics. He calls this doctrine `{\em 
Humean supervenience}'. It is stronger than {\em pointillisme} as defined in 
Section \ref{sssec;fourviol}, in that it is not relative to a theory. Roughly, it 
is relative to a possible world; (of course a metaphysician like Lewis who accepts 
the idea of a law of nature can link relativizations to a theory and to a possible 
world using the idea of the ``complete'' theory of a world, say as an 
axiomatization of all its laws of nature). And Lewis claims that it holds at the 
actual world.

The idea of Humean supervenience is that {\em all} truths supervene on truths 
about  matters of local particular fact: where `matters of local particular fact' 
is to be understood in terms of Lewis' metaphysics of natural properties, with the 
properties having spacetime points, or perhaps point-sized bits of matter, as 
instances. Thus he writes that Humean supervenience
\begin{quote} 
\ldots says that in a world like ours, the fundamental relations are exactly the 
spatiotemporal relations: distance relations, both spacelike and timelike, and 
perhaps also occupancy relations between point-sized things and spacetime points. 
And it says that in a world like ours, the fundamental properties are local 
qualities: perfectly natural intrinsic properties of points, or of point-sized 
occupants of points. Therefore it says that all else supervenes on the 
spatiotemporal arrangement of local qualities throughout all of history, past and 
present and future. (1994, pp. 225-226.)\footnote{Cf. also his (1986, pp. ix-x).} 
\end{quote}

Humean supervenience, so defined, is not widely believed---few philosophers sign 
up to all the notions deployed in its statement. But it has been a natural focus 
of metaphysicians' attention in the last twenty years, not least because Lewis has 
been the pre-eminent neo-Humean. In the literature, we can distinguish three broad 
groups of topics:\\
\indent (i) Issues about whether to analyse law, causation, chance etc., and 
``higher-level'' concepts about mind and language, in terms of the notions of 
Lewis' framework. Lewis (1986, p. xi-xiv) sketches how his work on all these 
topics provides a ``battle-plan'': i.e. roughly, a sequence of supervenience 
claims for these concepts.
More generally, much literature of neo-Humean stripe is concerned with 
how truths using familiar central concepts of common-sense knowledge and 
belief---concepts such as law, causation, the persistence of objects and mental 
and semantic concepts, such as belief and reference---might supervene on a basis 
acceptable to Humeans, though perhaps not exactly the basis proposed by Lewis. 
(For example: for laws, cf. Earman and Roberts (2006).) And some of these truths 
pose a challenge in that they seem {\em not} to thus supervene; cf. (iii) below.\\
\indent (ii) General metaphysical issues about the notions of Lewis' framework, in 
particular possible worlds and natural properties, and-or about related notions. 
For example, one well-known issue is: can possible worlds and natural properties 
be construed less ``realistically'' than Lewis proposes (e.g. Taylor 1993), and 
yet do the philosophical work they are meant to do?  More relevant to us will be 
Lewis' view of the intrinsic-extrinsic distinction, viz. that it can be analysed 
in terms of perfectly natural properties; (details in Section  
\ref{sssec;Bricker}). But as discussed in Sections \ref{sssec;ied} and 
\ref{sssec;2versions}, I can for the most part use only a much clearer 
distinction, viz. between the positive extrinsic properties and the rest, 
sub-divided in terms of temporal and spatial implications (or lack of them).\\
\indent (iii) Direct threats to Humean supervenience. There are two main examples. 
First, chance; which Lewis addresses in detail in (1986, pp. xiv-xvi, 121-131), 
and to his greater satisfaction in (1994). Second, persistence. For Lewis as a 
Humean wants to be perdurantist, as well as {\em pointilliste} in the sense of 
Humean supervenience: this means that he faces the rotating discs argument.

In this paper, I can set aside all of (i) and (iii), and all of (ii) except for 
the intrinsic-extrinsic distinction.

\subsection{Accommodating space and spacetime}\label{ssec;allowspace}
\subsubsection{An agreed concession}\label{sssec;qualifn}
As I mentioned, {\em pointillisme}'s need  to accommodate the structure of space 
and-or spacetime is agreed by all parties: in particular, by Lewis. In both the 
quotations above, Lewis includes relations of spatiotemporal distance (spacelike 
and timelike) in the supervenience basis. So his Humean supervenience is not so 
{\em pointilliste}, at least as regards the structure of space and/or spacetime, 
as it might at first seem.\\
\indent But no doubt even the most ardent {\em pointilliste} will find the 
inclusion sensible. That is: no one will hold that the structure of space and/or 
spacetime, in particular spatial and/or spatiotemporal metrical relations, is to 
supervene on intrinsic properties of points.\footnote{Here I  recall this paper's 
restriction to classical mechanics. So  I of course  set aside speculations in 
quantum gravity that classical spacetime structure emerges somehow from a 
``quantum pre-geometry'': speculations which, I agree, {\em might} have this 
structure emerge from (or even supervene on) intrinsic properties of some 
point-like objects. But I doubt it: most schemes for quantum pre-geometry are 
thoroughly ``relational'' rather than {\em pointilliste}. For surveys of such 
speculations, cf. e.g. Monk (1997), Butterfield and Isham (1999, Sections 1-4; 
2001, Section 5);  for a brief discussion  in relation to Lewis' Humean 
supervenience, cf. Oppy (2000, p. 88, 91-94).}\\
\indent The natural thing to say is, instead, that  the points and these metrical 
relations (and maybe also some spatiotemporal but {\em non}-metrical relations)  
form collectively a background, or canvas, on which other physical quantities 
taking various values get ``painted''. And it is to these latter that {\em 
pointillisme}'s doctrines are to apply. 

But there is no consensus (indeed, not much discussion) about how precisely to 
state the concession. More's the pity, since apart from the concession's own 
importance, it leads to the more general question (taken up in the following 
Subsections) how {\em pointillisme} can accommodate vectorial and tensorial 
properties.  

I begin with a preliminary issue. The concession obviously relates to the debate 
between relationist and substantivalist views of space and spacetime; and though I 
will not pursue this debate, I should register that this concession is not meant 
to prejudge it. There are two points here.\\
\indent (i): Though I spoke like a substantivalist, about a canvas of points, with 
various metrical and non-metrical relations between them, it is safe to assume 
that a relationist would appeal to similar relations  holding between items of 
their preferred ontology, i.e. bodies. (I set aside whether Leibniz's monads with 
only their intrinsic  properties might be enough to subvene all spatial and 
spatiotemporal facts!)\\
\indent (ii):  Similarly, my talk of a canvas of points, with metrical and 
non-metrical relations between them, was not meant to deny that the metric (or the 
other relations) could be dynamical, i.e. influenced by matter, in the way they 
are in general relativity.

So our question is what {\em exactly} metrical (and other geometrical) structures 
require. As physical  geometry has developed in the last two hundred years, these 
requirements have not only become subtler but have also become bound up with other 
properties and relations, especially of matter, in ways which threaten {\em 
pointillisme}. This issue will extend to Section \ref{ssec;allowvecs}. But let us 
start by raising the issues involved in as simple a context as possible: the 
length of a straight line in elementary geometry.

\subsubsection{The length of a line}\label{sssec;line}
The length of a line is the topic of a venerable paradox. The length of a straight 
line should surely be the sum of the lengths of a decomposition, i.e. an 
exhaustive set of mutually non-overlapping parts; and it seems legitimate to take 
as these parts the line's constituent points; but the length of each point is 
zero, and the sum of all these zeroes is presumably (though  a continuously large 
sum) zero---what else could it be? So the length of the line is zero! 

I stated this paradox in its most familiar form, as about summing lengths. But of 
course it can also be stated in philosophical  terms, as about supervenience: the 
length of a line surely does not supervene on the lengths of its points, on pain 
of being zero. That is no doubt why, as discussed in Section \ref{sssec;qualifn}, 
no {\em pointilliste} holds that lengths (or other metrical properties of lines, 
or indeed metrical relations between points) supervene on intrinsic properties of 
points.\footnote{There is of course a similar failure of supervenience for an 
extensive quantity such as mass, applied to a continuous body. For the point-sized 
parts of such a body have zero mass, so that the mass of the body is not the 
(uncountable) sum of the masses of those parts. Hawthorne notes this, while 
assessing how Lewis' metaphysics treats quantities (2006); and we will return to 
it in Section \ref{sssec;apreply}.}

This paradox is of course one of many that eventually prompted the development of 
measure theory. And as noted in (2i) and (c) of Section \ref{Intr}, measure theory 
invites philosophical scrutiny because (i) it has some well-nigh paradoxical 
results of its own, like the Banach-Tarski paradox, and (ii) it is connected to 
mereology. But I shall not need details about these topics. I only need to 
present:\\
\indent (a): a philosophical reply to this paradox; though it does not block the 
paradox, it introduces an important metaphysical trichotomy among relations; \\
\indent (b): the main idea of the technical measure-theoretic reply to this 
paradox.   

\paragraph{3.3.2.A The philosophical reply}\label{332APhilReply} The philosophical 
reply is just that length is a property of the line as a whole, where `line as a 
whole' can be taken to mean either the set, or the mereological fusion, of its 
points (or of its extended parts). That is, the length does not supervene on the 
properties of the points (or other parts). This is surely true, so far as it goes. 
But it is not enough to block the paradox, since it does not pinpoint what is 
wrong with the premise that the length of a straight line is the sum of the 
lengths of an exhaustive set of mutually non-overlapping parts. (The technical 
reply will do this.)

However, this reply prompts  a trichotomy among relations corresponding to the 
intrinsic-extrinsic dichotomy among properties: a trichotomy that will be useful 
in what follows. Lewis  states the trichotomy clearly (1983a, p. 26 fn. 16; 1986a, 
p. 62); and I shall adopt his proposed terminology (which has become widespread). 
Though he explains it in terms of his preferred understanding of intrinsic and 
extrinsic properties (viz. defined in terms of his natural properties), the 
trichotomy can be explained in the very same words, using other understandings of 
intrinsic and extrinsic. In particular, it can be thus explained using Section 
\ref{sssec;ied}'s suggested understanding of `extrinsic' as `positive extrinsic' 
and `intrinsic' as `not positive extrinsic'; (or using Lewis' ``second favourite'' 
analysis developed by Langton and him (1998, p. 129)). The trichotomy also uses 
the idea of the mereological fusion, or composite, of objects: an idea I am happy 
to accept, and for which there is a powerful argument (Lewis 1986a, pp. 212-213, 
developed by Sider 2001, pp. 121-139).

\indent (1) An {\em internal} relation is determined by the intrinsic properties 
of its relata. So if $xRy$, and $x'$ matches $x$ in all intrinsic properties, and 
$y'$ matches $y$ in all intrinsic properties, then we must have $x'Ry'$. So any 
relation of similarity or difference in intrinsic respects is internal; for 
example, if height is an intrinsic property, then `being taller than' is an 
internal relation.

\indent (2) On the other hand, there are relations, notably relations of 
spatiotemporal distance, that are not internal, but {\em do} supervene on the 
intrinsic nature of the composite (mereological fusion) of the relata. Thus 
suppose $x,y$ are point-particles 1 metre apart. Then it seems reasonable to say 
both of the following:---\\
\indent\indent (i): There could be point-particles $x',y'$ that intrinsically 
match $x$ and $y$ respectively, and that are 2 metres apart---so that distance is 
not internal. But on the other hand:\\
\indent\indent (ii): Any object intrinsically matching the {\em fusion} or {\em 
composite} of $x$ and $y$ would have two parts intrinsically matching $x$ and $y$, 
1 metre apart.\\
\indent\indent Accordingly, Lewis calls relations that supervene on the intrinsic 
nature of the fusion of the relata, {\em external}; and he takes (ii) to show that 
spatiotemporal relations are external.

\indent (3) Finally, there are relations that do not supervene even on the 
intrinsic  nature of the composite of the relata; i.e. relations that are neither 
internal nor external. Lewis' example is the relation {\em having the same owner}: 
$x$ and $y$ could intrinsically match $x'$ and $y'$ respectively, and their 
composites might also match; and yet $x$ and $y$ might have the same owner, while  
$x'$ and $y'$ do not. But more relevant to us than ownership, geometry and 
mechanics provide many examples of such relations. The objects $x,y,x',y'$ could 
be solid bodies, again with each pair, $x,x'$ and $y,y'$, intrinsically matching, 
and the composites $x+y, x'+y'$ also matching---and yet the centre of mass of $x$ 
and $y$ might be a certain distance from a body of some kind $Z$, while the centre 
of mass of $x'$ and $y'$ is not.

\paragraph{3.3.2.B The technical reply}\label{332BTechnlReply}
The technical reply to the paradox of length comes  from measure theory. It blocks 
the paradox by denying the premise that the length of a straight line is the sum 
of the lengths of any decomposition (exhaustive set of mutually  non-overlapping 
parts) of the line. It upholds this only for certain decompositions. The main idea 
is to consider only decompositions containing points and intervals, and to accept 
the additivity of length for at most denumerably large sums. These ideas give a 
rich theory which can be extended to cover  area and volume, as well as length; 
and which underpins the theory of integration.

But we do not need further details of measure theory. For us the point is that, 
even in the elementary geometry of Euclidean space ($\mathR, \mathR^2$ etc.), we 
cannot say all that is true in terms just of intrinsic properties of points. For 
we need to assign lengths to spatial intervals. And---to use Lewis' terms---the 
length of an interval is surely not an internal relation between the interval's 
end-points, since any two points seem to match in intrinsic properties. Besides, 
the length of an interval does not seem to be an internal relation  between all 
the interval's uncountably many points, since: (i) the points of any two intervals 
seem to match pairwise in intrinsic properties, and (ii)  additivity of length 
fails for an uncountable set.

 On the other hand, it seems reasonable to say, as Lewis does, that:\\
\indent (a): intervals are composites or fusions of their points; and \\
\indent (b): intervals matching in their intrinsic properties are congruent; (cf. 
(2) (ii) above).\\
If we say (a) and (b), then it follows that the length of a straight interval is 
an external relation among the interval's points. (So far, this is a relation of 
uncountable polyadicity: the next Subsection will ask  whether the relation can be 
taken to be just a dyadic relation  between the interval's two end-points.)

To sum up: to describe length, even the length of a straight interval in euclidean 
geometry, {\em pointillisme} must concede that  it needs to go beyond intrinsic 
properties of points, and even relations that are internal in Lewis' sense. But so 
far, it seems (cf. (2)(ii) in Section \ref{sssec;line}.A, and (b) above) it can 
manage with what Lewis calls external relations.\footnote{Incidentally, returning 
to the original paradox of length: the practice of measure theory seems 
indifferent between the following options (and surely metaphysicians can be as 
well):\\
\indent a) to allow {\em ab initio} that each point has a length, viz. zero; and 
avoid paradox by denying  uncountable additivity of length;\\
\indent b) to ascribe length primarily only to sets of points in a certain 
well-behaved family of sets.\\
But the technical need for the family to have certain closure properties is likely 
to lead to singleton sets of points being included---as in the usual choice of 
family, the Borel sets.}

\subsubsection{Accommodating more geometry}\label{sssec;moregeomy}
The further development of geometry, including more general geometries than the 
Euclidean line or plane, reinforces the point that we cannot say all that is true 
in terms just of intrinsic properties of points. But as we shall see, it is 
doubtful that we can manage with just Lewis' category of external relations. That 
is, it is doubtful that all {\em pointillisme} needs to do, in order to 
accommodate spatial and spacetime structure, is to admit the network of external 
relations of spatial and spatiotemporal distance. 

Talk of `spatial and spatiotemporal relations' tends to suggest that space or 
spacetime is a {\em metric space}, in the usual mathematical sense that (given a 
unit of length) there is a real-valued function on pairs of points: to any pair of 
points $x,y$ is assigned a distance $d(x,y) \in \mathR$. (So each real number 
determines a binary relation on points; and for a relativistic spacetime, $d$ need 
not be positive-definite.) But the development of geometry  and physics has shown 
that this is {\em much} too limited a conception of spatial (or spatiotemporal) 
structure. One needs to distinguish various subtly related levels of structure: 
for example, geometers distinguish topological structure,  differential structure, 
metrical structure, and many more. Besides, most of these kinds of structure 
include definitions of irreducibly global features of the space concerned; for 
example, a space can have the global topological feature of being simply 
connected, i.e. such that all closed curves can be continuously deformed to a 
point.

But here I will focus only on ``local'' metrical structure.\footnote{I put `local' 
in scare-quotes, since I here intend a vaguer meaning than that of (iii) of 
Section \ref{sssec;3maths}. But in what follows, the meaning will always be clear 
from the context. Note also that though I will discuss only space, all I say 
carries over intact to spacetime.} Even without developing the formal details 
(which go back to Gauss and Riemann), we will be able to see that geometry (and 
more generally physics) needs to attribute to points both vectorial and tensorial 
properties---raising the question, pursued in the next Subsection, whether {\em 
pointillisme} can accommodate such properties.

Gauss and Riemann proposed that we take as the primary notion, the length of a 
curve  between two points; (so since a pair of points is in general connected by 
infinitely many curves, any such pair is associated with infinitely many lengths, 
not just one). This proposal is adopted by modern spacetime theories, in 
particular by the most successful such theory, general relativity. So an advocate 
of {\em pointillisme} (or some similar doctrine of local supervenience, such as  
Humean supervenience)  would do well to formulate their doctrine so as to 
incorporate, or at least allow for, this proposal. 

At first sight, it seems that the {\em pointilliste} can manage just fine. She 
only needs to apply Lewis' idea of external relations (or perhaps, some similar 
notion), not to the end-points of a straight interval or to all the points of a 
straight interval, but to all the points of an arbitrary curve. Thus she can take 
an arbitrary curve  as the fusion of its points, and the length of the curve as an 
external relation, albeit of uncountable polyadicity, among the points: an 
external relation which determines an intrinsic property of the fusion. I presume 
that Lewis, who was well aware of the Gauss-Riemann conception of geometry, would 
have said this. That is, he would have taken `spatiotemporal distance relations' 
in his definition of Humean supervenience to allow for this conception---and not 
to be committed to the idea of a metric space.\footnote{But Lewis seems never to 
have pursued the question exactly what relations he should propose as fundamental 
for modern geometry and topology; and we shall shortly see trouble for him, as for 
other {\em pointillistes}.} 

Besides, the structure required to define the lengths of all curves  is given 
``locally'' in a way that at first sight seems congenial to a {\em pointilliste}. 
In particular, it seems that  the {\em pointilliste} does not need to postulate 
continuously many external relations, each of uncountable polyadicity, one for 
each congruence-class of curves. For:\\
\indent (i): measure theory can be extended to apply to this kind of geometry, so 
that the length of a curve is the sum of the lengths of a countable decomposition 
of it;\\
\indent (ii): similarly, calculus can be extended to apply to this kind of 
geometry, so that the length of a curve is given by an integral along 
it---intuitively, an uncountable sum of infinitesimal contributions one for each 
infinitesimal element of the curve.\\
\indent These features, (i) and (ii), reflect the fact that the Gauss-Riemann 
conception of metrical structure presupposes topological and differential 
structure: which make sense, respectively, of the notions of continuous function, 
and differentiable function. 

But there is a devil in the details. The details of how to define the length of a 
curve require us to attribute vectors and tensors to a point. A bit more 
precisely: we need to attribute:\\
\indent (i) to any point on any curve, the {\em tangent vector} to the curve at 
that point; and so\\
\indent (ii) to any point, the set of all such tangent vectors at it (which form a 
vector space, called the {\em tangent  space});\\
\indent (iii) to any point, a {\em metric tensor} which maps pairs of tangent 
vectors at the point to real numbers---generalizing the elementary scalar product 
of two vectors. (More details about (i)-(iii) in the next Subsection.)\\
\indent Thus the {\em pointilliste} has to face---already in geometry, even before 
considering physics' description of matter--- the question announced in Section 
\ref{ssec;pismprospect}: can a property represented by a vector or a tensor be 
intrinsic to a point?

\section{Accommodating tangent vectors and the metric 
tensor}\label{ssec;allowvecs}
\subsection{Bricker and others}\label{sssec;focusBric}
As I announced in Section \ref{ssec;pismprospect}, my main effort in this Section 
will be to report and criticise Bricker's (1993) discussion of this question, for 
metrical properties. I choose him for two reasons.\\
\indent (i): His paper is an unusually thorough and perceptive attempt to relate 
vectors and tensors, as they are treated in modern geometry, to the modern 
metaphysics of properties. So it repays detailed scrutiny.\\
\indent (ii): His paper illustrates the tendency that, as I said at the end of 
Section \ref{ssec;pismprospect}, I want to reject: the tendency of some 
contemporary metaphysicians to reconcile {\em pointillisme} with physical 
theories' use of vectorial properties, which seem extrinsic to points, by 
proposing some heterodox construal of the properties in question. Bricker proposes 
that in order to understand space's or spacetime's metrical structure as 
intrinsic, we should appeal to non-standard analysis. I will deny this heterodoxy: 
instead, we can and should reject {\em pointillisme}.

But before going into the details of Bricker's discussion, I should register that 
other metaphysicians  have also addressed our question; though (so far as I know) 
more briefly and with less attention to technicalities than Bricker. (Besides, 
they are not all attracted by {\em pointillisme}, or by the above tendency.)

For example, Robinson maintains that the directionality of a vector forbids it 
from representing an intrinsic property: `direction seems to me an inherently 
relational matter' (1989, p. 408). And he would presumably say the same about 
tensorial properties. (His paper is about the rotating discs argument; I discuss 
its proposals in 2006.) Robinson gives an argument for this, using Lewis' notion 
of {\em duplicates}, i.e. objects that share all their intrinsic properties. He 
also credits Lewis for the argument; so presumably  Lewis himself  thought at the 
time (ca. 1988) that vectorial properties could not be intrinsic. The argument 
combines two intuitions:\\
\indent (a): It seems that a vectorial property  could not be instantiated in a 
zero-dimensional world consisting of a single point; though since arbitrarily 
close points define a direction, there is of course no lower limit to the ``size'' 
of a world in which a point instantiates a vectorial property. \\
\indent (b): But it also seems that, since a point in an extended world that 
instantiates a vectorial property is indeed a {\em point}, it could have a 
duplicate that existed on its own, i.e. was the only object in its world.\\
\indent Taken together, (a) and (b) imply that duplicate points might differ in 
their vectorial  properties; so that any such property is not 
intrinsic.\footnote{Other metaphysicians also maintain that vectorial properties 
are extrinsic to points. For example, Black (2000, p.103) holds that vectorial 
properties can be intrinsic only for the special case of vectors on a manifold 
with a flat connection; i.e. roughly, a manifold in which there is a unique 
preferred way to compare vectors located at different points. (His discussion is 
briefer than Robinson's; since the topic is less relevant  to his paper's main 
aims, than it is to Robinson's.) And Zimmerman (1998, p. 277-278) and Oppy (2000, 
pp. 79-82) are similarly inclined; though they also discuss sympathetically the 
opposing view.} 

But for anyone who is attracted by {\em pointillisme}, and is aware of physical  
theories' use of vectorial properties, this is a very uncomfortable conclusion. 
Lewis himself is a case in point. Indeed, he seems to have come round to believing 
that vectors can represent intrinsic properties of points, sometime between ca. 
1988 and ca. 1993. For in a discussion of Humean supervenience (1994, p. 474), he 
says he is inclined to think that vectorial properties are, or at least can be, 
intrinsic: `any attempt to reconstrue them as relational properties seems 
seriously artificial'. But, so far as I know, that is all Lewis says by way of 
defending the idea; (though in his (1999) he used the idea to try and reply to the 
rotating disc argument---unsuccessfully I maintain (2006)). In any case, I now 
turn to Bricker's extended struggle to avoid the uncomfortable conclusion.

\subsection{Bricker's three claims about metrical structure}\label{sssec;Bricker}
\subsubsection{Bricker's metaphysical framework}\label{342ABrickerFwk} 
Bricker's (1993) overall aim is metaphysical understanding of spatial (or 
spatiotemporal) relations. He adopts a metaphysical framework very close to 
Lewis'---with of course all due acknowledgement (1993, pp. 273-5). The ingredients 
we need are:---\\
\indent (i) He speaks of possible worlds and perfectly natural properties and 
relations. He applies mereology freely to points of space and spacetime; (in fact, 
substantivalism about space and spacetime is widespread among  analytic 
metaphysicians). And so he takes worlds and parts of worlds as {\em possibilia}. 
\\
\indent (ii) He says that any two {\em possibilia} $X,Y$ are {\em duplicates} iff 
there is a one-to-one correspondence between their parts that preserves all 
perfectly natural properties and relations. He calls any such correspondence an 
$(X,Y)$ {\em counterpart relation}, and corresponding parts are 
$(X,Y)$-counterparts of each other. (So in this Subsection, `counterpart' is tied 
to `duplicate' and so will not have the usual Lewisian connotations of allowing 
vagueness and extrinsicness.)\\
\indent (iii) He says that a property is {\em intrinsic} iff any two duplicates 
both have it or both lack it. (Otherwise the property is {\em extrinsic}.) It 
follows that:\\
\indent \indent (a) the class of all {\em possibilia} is partitioned by the 
equivalence relation of being  duplicates; and\\
\indent \indent (b) an intrinsic property corresponds to a union of cells of this 
partition; and\\
\indent \indent (c) all perfectly natural properties are intrinsic.\\
\indent He extends the notion of intrinsic to relations by saying that a relation 
is intrinsic iff it is either internal or external in the senses of Lewis (defined 
in Section \ref{sssec;line}.A); otherwise the relation is extrinsic.\\
\indent (iv) He assumes (following Lewis 1986a, pp. 86-92) a principle of 
recombination for spatial or spacetime points. This is a principle of modal 
plenitude, inspired by a Humean denial of necessary connections between distinct 
existences: ``anything can follow anything''. Stated for spatial points, it holds: 
for any points $p$ and $q$, perhaps from spaces of different worlds, there is a 
world whose space is a duplicate of the space of $p$, except that it contains a 
duplicate of $q$ where the duplicate of $p$ would be (1993, p. 290); and similarly 
for spacetime.

I do not  endorse this framework. But in discussing Bricker, I will use it (and a 
variant of it considered by him). Though it would be a good project to ascertain 
how well Bricker's arguments fare under a different framework (in particular, 
under weaker assumptions about the intrinsic-extrinsic distinction), it is not a 
project for this paper. Here it must suffice to note that if we used my 
distinction between positive extrinsics and the rest, advocated in Section 
\ref{sssec;ied}, the main points of my critique of Bricker below, would carry over 
intact. But I shall not spell this out point by point, from now on. I just note 
here that:\\
\indent (a): Since my distinction takes `intrinsic' to mean `not positively 
extrinsic', it yields more intrinsic properties than does Bricker's (or Lewis') 
framework; and so a logically stronger notion of duplicatehood as sharing of all 
intrinsic properties. \\
\indent (b): Bricker's argument for the spacetime metric being extrinsic to points 
(Section \ref{sssec;metrextr}) remains valid on my distinction's construal of 
`extrinsic' as `positive extrinsic'. For Bricker's argument implicitly appeals to 
positive extrinsicality.  \\
\indent (c): My anti-{\em pointilliste} reply to Bricker (Section 
\ref{sssec;apreply}) is unaffected by adopting my distinction.

Bricker goes on to connect his framework with the Gauss-Riemann conception of 
distance, as endorsed by general relativity. His discussion includes aspects 
(1993, pp. 275-286) which we can skip, in particular: (a) a comparison with two 
other conceptions of distance (which he dubs the `naive' and `intrinsic' 
conceptions); and (b) a discussion of how the principle of contact-action (denial 
of action-at-a-distance) bears on the the Gauss-Riemann conception. Setting these 
aside, I read Bricker as connecting his metaphysical framework with the 
Gauss-Riemann conception, as follows. He assumes that:\\
\indent (i) the perfectly natural properties and relations, that are instantiated 
at a possible  world that has laws of nature, figure in that world's  laws 
(however the notion of a law of nature is to be analysed);\\
\indent (ii) general relativity is a logically possible theory, giving the 
gravitational and metrical laws of some possible worlds;\\
\indent (iii) general relativity can be ``formulated locally''; which is taken to 
imply, as regards metrical structure, formulated in terms of local metrical 
relations. \\
\indent Taking these assumptions together, he concludes that the property of 
having such-and-such a local metric tensor  is a perfectly natural property, and 
is instantiated at points in general relativistic worlds.

So far, Bricker is in a position like the one we articulated at the end of Section 
\ref{ssec;allowspace}, in which the {\em pointilliste} seemed well able to manage 
local metrical structure. Bricker also notes (as we did) that since even topology 
brings in irreducibly global properties of space like being simply connected, 
there can be no sweeping supervenience of the global on the local. So he 
formulates a doctrine he calls {\em Einsteinian supervenience}, on analogy with 
Lewis' Humean supervenience: there is `a manifold of spacetime points ... and a 
distribution of perfectly natural local properties (including local metrical 
properties) over those points; all else supervenes on that' (1993, p. 288).

\subsubsection{Bricker's three claims}\label{342BBricker3Claims} 
Bricker then notices what I called `the devil in the details', i.e. the fact that 
local metrical structure attributes vectorial and tensorial properties to points; 
and he goes on to address the question whether such properties are intrinsic, in  
terms of his metaphysical framework (1993, pp. 288f.). He argues for the following 
three claims (in order, with my added mnemonic labels).

\indent (MetrExtr): The metric tensor, as standardly conceived in differential 
geometry, represents an {\em extrinsic} property of a point.

\indent (VetoExtr): The obvious (and anti-{\em pointilliste}) response to the 
conflict between this and the metric being perfectly natural---viz. that some but 
not all perfectly natural properties are intrinsic---does not work. For, Bricker 
argues, it clashes with the Humean principle of recombination for spacetime 
points. That is, Bricker rejects this response as engendering  necessary 
connections between distinct existences, viz. a point and its surrounding space.\\
\indent So Bricker claims we do better to revise our conception of the metric 
tensor, as follows.

\indent (Heterodox):  We should take the metric to represent an intrinsic property 
of an {\em infinitesimal neighbourhood} of a point. Bricker cites Robinson's 
non-standard analysis as justifying taking such neighbourhoods as genuine 
mathematical objects, rather than as a {\em facon de parler} for calculus' 
standard notion of limit as ``$\forall\exists\forall$'' (e.g. for a real sequence 
$\{a_n\}$: $\forall \; \varepsilon > 0 \;\; \exists N \; \; \forall m,n > N \;\; 
\mid a_m - a_n \mid < \varepsilon $).

\indent So Bricker's overall conclusion is radical: that in order to save {\em 
pointillisme}, we should revise the foundations  of differential geometry. In the 
next three Subsections, I will report his arguments for (MetrExtr) to (Heterodox). 
Then in the last Subsection (Section \ref{sssec;apreply}), I will deny his 
conclusion. Since I hold no brief for {\em pointillisme}, I see no reason to pay 
his price of revising the foundations of differential geometry.

\subsection{The standard metric is extrinsic}\label{sssec;metrextr}
 Bricker's argument for (MetrExtr)---{\em the metric tensor, as standardly 
conceived, represents an extrinsic property of a point}---is not absolutely 
precise. But it uses more technicalities about local metrical structure than I 
have introduced so far, in particular  differential geometry's idea that the 
tangent  vectors at a point be taken to be directional derivative operators. So I 
need to review this; I shall give rather more detail than Bricker does. 

\indent (i): First, the set of spatial or spacetime  points is assumed to form a 
manifold $\cal M$. The definition of `manifold' is elaborate, and was only given 
in its modern formal guise in the 1930s---and fortunately I can skip it! It 
suffices to say that the definition gives sense to various crucial ideas such as 
the dimension of a manifold, its boundary (if any), its global topological 
structure (e.g. being simply connected), the idea of a smooth scalar function i.e. 
a smooth real-valued function defined on a subset of the manifold---and most 
important for us, the idea of a smooth curve in the manifold, which is taken as a 
map $q$ from an interval of real numbers $I \subset \mathR$ to $\cal M$. (Here 
`smooth' refers to differentiability a specified number of times.) As I said in 
Section \ref{ssec;allowspace}, the {\em pointilliste} will be hard pressed to 
account for this manifold structure: but I  will not labour this point.  \\
\indent (ii): Any curve $q$ thus includes in its definition its real-number 
parameter, $\lambda$ say. So, understanding the tangent vector to the curve at the 
point $q(t), t \in I$, in an intuitive way: the tangent vector  specifies a 
directional derivative of any scalar function $f$ defined on a neighbourhood, $N$ 
say, of the point $q(t)$, $f:N \subset{\cal M} \rightarrow \mathR$. (For the 
direction of the curve at $q(t)$, together with  the ``rate at which $\lambda$ 
ticks away'', defines an ``instantaneous rate of change'' of $f$.)\\
\indent (iii) It is convenient to {\em identify} the tangent vector to the curve 
$q$ at the point $q(t)$  with the directional derivative operator acting on the 
set, $\cal R$ say, of all scalar functions defined on some neighbourhood of the 
point $q(t)$: $\frac{d}{d{\lambda}}\mid_{q(t)}: f \in {\cal R} \mapsto 
\frac{df}{d{\lambda}}\mid_{q(t)} \in \mathR$.\\
\indent \indent Why is it convenient? In short: because the directional derivative 
operators behave just like tangent vectors. For example, for an $n$-dimensional 
manifold $\cal M$, the directional derivative operators at any point $p \in {\cal 
M}$ form an $n$-dimensional vector space, just as one would want the  tangent 
vectors to do: think of the 2-dimensional tangent plane at a point $p$ on the 
surface of a sphere. This vector space is called the {\em tangent space} at $p, 
T_p$.\\
\indent \indent (Other equivalent  identifications are also used: some 
presentations identify a tangent vector at $p \in {\cal M}$ with an equivalence 
class of curves through $p$---intuitively, curves that are all tangent to each 
other at $p$ and with parameters ``ticking'' at the same rate.)   \\
\indent (iv): To define the length of a curve requires still further structure: 
structure which is not fixed by the postulation of a manifold, with all its 
tangent vectors $V \in T_p$ at each point $p$. Namely, it requires a metric tensor 
$g$, which is an assignment to each point in $p \in {\cal M}$ of a mapping from 
pairs of vectors $\langle U,V \rangle$ with $U,V \in T_p$ to $\mathR$: a mapping 
of a certain sort that generalizes the elementary scalar product of vectors. So 
$g: \langle U,V \rangle \mapsto g(\langle U,V \rangle) \in \mathR$. This metric 
tensor applied to the pair $\langle V,V \rangle$, where $V$ is the tangent vector 
to a curve $q$ passing through $p$, gives in effect the squared length of the 
``infinitesimal part'' of the curve at $p$. Now, if we let $p$ vary from one point 
of the curve to another and add up the corresponding contributions, we are  
performing an integration. So integrating (the square-root of) $g(\langle V,V 
\rangle)$  gives the length of the curve. One can prove that (as one would want) 
the length of a curve depends on the metric tensor used, but not on how the curve  
is parameterized.    

To connect (i)-(iv) with Bricker's claim (MetrExtr), one needs some  
``bridge-principles'' between the mathematical constructions and philosophical  
notions such as that of an intrinsic property. For this, Bricker proceeds as 
follows. He defines (1993, p. 289) a property $P$ of points to be {\em local}  iff 
for any points $p,q$, any neighbourhood $N$ of $p$ and any neighbourhood $M$ of 
$q$:
\begin{quote} 
if $N$ is a duplicate of $M$, and $p$ is an $(N,M)$-counterpart of $q$, then $P$ 
holds either of both $p$ and $q$ or of neither (i.e. $p$ and $q$ match as regards 
$P$).
\end{quote}
So, roughly speaking, Bricker calls a property $P$ of points `local' if whether a 
point $p$ possesses $P$ is wholly determined by the intrinsic nature of any 
arbitrarily small neighbourhood of $p$. So, {\em modulo} the use of metaphysical 
ideas of intrinsic property, duplicatehood etc., this usage clearly corresponds to 
mathematicians' use of `local' (cf. (iii) in Section \ref{sssec;3maths}). 
  
It follows that for Bricker any intrinsic property of points is local, since 
counterpart points, being duplicates of each other, share all their intrinsic 
properties. But, Bricker maintains, the converse fails: there are local but 
extrinsic properties of points. These he dubs {\em neighbourhood-dependent}.

He briefly discusses as examples from elementary calculus, derivatives of 
functions, in particular instantaneous velocity. He says the instantaneous 
velocity of a point-particle at position $x$ at time $t$, i.e. at a spacetime 
point $p$, depends on where the particle is at other times; and so is a 
neighbourhood-dependent, but not intrinsic, property of $p$. The `so' here is not 
spelt out precisely, i.e. by justifying the implicit premise about duplicate 
spacetime regions containing the particle (or its counterpart). But Bricker's 
intuition is  clear enough: as we emphasised already in Section \ref{sssec;3maths} 
and \ref{ssec;moredetail}.B, instantaneous velocity and momentum are temporally 
extrinsic since for example they imply the object's existence at other times. 
Besides, the intuition is shared by others---as we will see when we return to 
instantaneous velocity in the next Section.

Bricker goes on to claim by analogy that in differential geometry all the tangent 
vectors at a point $p$ `give information not just about $p$, but about the space 
immediately surrounding $p$ ... in short ... neighbourhood-dependent information 
about $p$'. To which he adds: `since the local metric at $p$ is an operator on 
tangent vectors, it inherits neighbourhood-dependence from its operands' (1993, p. 
289). 

Again, Bricker's argument here is not entirely precise. He cannot really prove 
that {\em any} property represented by an element of tangent space is extrinsic; 
for his metaphysical apparatus does not tie its notions of perfectly natural 
property, and so duplicate, and so $(X,Y)$-counterparthood, sufficiently tightly 
to the notions of differential geometry. A footnote admits that (as in my (iii) 
above), tangent vectors are directional derivative  operators; but again there is 
no justification for the implicit premise about duplicate spacetime regions. But 
fair enough, I say: his intuition is again both clear and shared by others.

And the intuition is enough to deliver Bricker his problem. That is: the metric's 
being neighbourhood-dependent contradicts the previous claim that it is perfectly 
natural (i.e. perfectly natural because mentioned in the laws of general  
relativity)---once we recall that according to his metaphysical framework, all 
perfectly natural properties  are intrinsic.

 In response, Bricker considers two tactics for escape from contradiction: an 
obvious one which he rejects in the second stage of his argument (Section 
\ref{sssec;natextrinsics?}); and an unobvious one which he endorses in the last 
stage (Section \ref{sssec;heterodox}).

\subsection{Vetoing perfectly natural extrinsics}\label{sssec;natextrinsics?}
Bricker now argues for:\\
\indent\indent (VetoExtr): {\em The obvious anti-pointilliste  response to the 
contradiction between the metric being neighbourhood-dependent and perfectly 
natural---viz. that some but not all perfectly natural properties are 
intrinsic---clashes with the Humean principle of recombination for spacetime 
points. That is: it engenders  necessary connections between a point and its 
surrounding space.}
 
Bricker first considers saying that some but not all perfectly natural properties 
are intrinsic. So the idea is that the perfectly natural but extrinsic  properties 
of points  include vectorial and tensorial properties, like having a metric tensor 
with such and such features.

 Bricker notes that this response implies that his previous definition of 
`duplicate' bifurcates into a weaker and a stronger notion. The weaker notion is 
that of {\em
 intrinsic duplicates}: this requires only that the one-one correspondence between 
the parts of objects $X$ and $Y$ preserve the {\em intrinsic} perfectly natural 
properties and relations. (Recall that Bricker calls a relation `intrinsic' iff it 
is internal or external in Lewis' sense, given in Section \ref{ssec;allowspace}.) 
The
 stronger notion, which Bricker calls  {\em local duplicates}, has the same 
definition, word for word, as the previous definition of duplicates: $X$ and $Y$ 
are local duplicates iff there is a one-one correspondence between their parts 
preserving  {\em all} perfectly natural properties and relations. Bricker proposes 
that we now define a local property as one that never differs between local 
duplicates. So it is now built in to the definitions 
that perfectly natural properties are local---just as previously it was built in 
that they were intrinsic. Returning to geometry, the idea will be that such 
perfectly natural, and so local, properties include vectorial and tensorial 
properties, like having a metric tensor with such and such features.  

So far, so good. But there is a clash with Bricker's Humean principle of 
recombination for  points ((iv) of Section \ref{342ABrickerFwk}): that for any 
points $p$ and $q$, there is a world whose space is a duplicate of the space of 
$p$, except that it contains a duplicate of $q$ where the duplicate of $p$ would 
be. More precisely: Bricker says there is a dilemma. For this principle must now 
refer either to (A) local duplicates, or to (B) intrinsic duplicates: and on 
either interpretation, Bricker sees trouble. I will reply that the second 
interpretation, (B), is fine---provided one is not a {\em pointilliste}.  

\subsubsection{Trouble with local duplicates}\label{344A;vslocaldupls}  
If  the principle of recombination refers to local duplicates, then it  will yield 
contradictory worlds when $p$ and $q$ have contrary perfectly natural, extrinsic 
(but of course local) properties. Bricker gives as his example positive and 
negative curvature. He writes: `suppose that $p$ is surrounded by positively 
curved space, $q$ by negatively curved space. Then a world whose space is a 
duplicate of the space of $p$ but with a {\em local} duplicate of $q$ in $p$'s 
place must be both positively and negatively curved in the immediate neighbourhood 
of $q$' (1993, p. 290; the last phrase of course means `immediate neighbourhood of 
the duplicate of $q$').

 Here I should amplify  Bricker's example---and point out a problem raised by it. 
Given a metric, one can define a {\em scalar} function, in the usual mathematical 
sense of `scalar' (viz. a function from the manifold $\cal M$ to $\mathR$, so that 
its value at a point $p \in {\cal M}$ is the same, independently of any choice of 
coordinate system), called the {\em scalar curvature} $R$, that has the following 
remarkable property: although it is a scalar, at each point $p$ its value $R(p)$ 
is a numerical measure of how curved is the geometry in a neighbourhood of $p$. 
(In fact a metric is sufficient  but not necessary to define $R$: a connection 
also allows one to  define scalar curvature.) So Bricker is no doubt here assuming 
that:\\
\indent (i): $p$ and $q$ have positive and negative scalar curvature, 
respectively, i.e. $R(p) > 0$ and $R(q) < 0$; (and if we like, we can take him to 
assume that all points in their respective neighbourhoods have positive and 
negative scalar curvature);\\
\indent (ii): the scalar curvature $R$ is perfectly natural but extrinsic: (more 
precisely, it is a determinable whose determinates, given by specific values 
$R(\;)= 5$ etc., are perfectly natural but extrinsic).

 Assumption (ii) raises a problem. Hitherto, we have implicitly assumed that 
scalar functions represent intrinsic properties of points: our worries have  
concerned only vectorial and tensorial properties. Now we see there is also a gap 
between:\\
\indent (a): the mathematical notion of a scalar, which is a matter of how a 
quantity transforms (viz. trivially: it takes the same value in all coordinate 
systems); and\\
\indent  (b): the metaphysical idea of intrinsicness.\\
That is: some scalars can `give information not just about $p$, but about the 
space immediately surrounding $p$'---to quote Bricker's words from his discussion 
of tangent vectors (quoted in Section \ref{sssec;metrextr}'s discussion of 
Bricker's (MetrExtr)).\\
\indent So Bricker owes us a discussion of how exactly being a scalar, and being 
intrinsic, relate. But this is not to say that the onus is only on Bricker. So far 
as I know, this is a lacuna in the whole metaphysical literature. (Cf. comment (a) 
at the end of Section \ref{sssec;3maths}.)

\indent To sum up: the metaphysical literature assumes that any scalar represents 
an intrinsic property of points, so that the {\em pointilliste} need ``only'' 
worry whether vectors, tensors etc. do as well. But now we see that {\em 
pointillistes} should also worry about scalars such as the scalar curvature. 

\subsubsection{Alleged trouble with intrinsic  duplicates}\label{344B;vsintrdupls} 
On the other hand, suppose the principle of recombination refers to intrinsic  
duplicates. Then contradictory worlds are avoided; but, says Bricker, the 
principle is now too weak to capture the spirit of the Humean denial of necessary 
connections between distinct existences. For the principle now rules out necessary 
connections between the {\em intrinsic} natures of distinct objects. But on the 
present response, an object's ``nature'' can include more than its intrinsic 
nature, viz. its perfectly natural extrinsic properties. So, says Bricker, the 
principle's free combinability of intrinsic natures is not enough to prevent 
unwanted necessary connections. 

I reply that this second horn of Bricker's dilemma has force only for a {\em 
pointilliste}. To see the point, let us take an example.  Bricker does not give 
one: but he could add to the above example of $p$ and $q$, as follows. If one 
scalar function, say temperature $\theta$, represents an intrinsic property, while 
the scalar curvature $R$ represents  a perfectly natural extrinsic property (and 
for simplicity, there are no other scalar, vector or tensor functions to 
consider), the principle of recombination yields a world that has in $p$'s place 
an intrinsic duplicate of $q$---i.e. a point with:\\
\indent (a) the same temperature $\theta$ that $q$ has (in its world), but\\
\indent (b) the scalar curvature $R$ that $p$ has (in its world), i.e. a positive 
value, not $q$'s negative value.\\
\indent  In short, the fact that the neighbourhood of this point is a duplicate of 
$p$'s neighbourhood forces the duplicate of $q$ going into $p$'s place to ``shed'' 
its negative scalar curvature.

I take it that Bricker would see this  as an unacceptable un-Humean necessary 
connection between the point $p$ and its surrounding space. But I claim that the 
wise Humean has no worries here: the necessary  connection merely reflects the 
extrinsicality of scalar curvature, so that the value of the scalar curvature in 
the surrounding space can constrain its value in $p$'s place.\footnote{A 
mollifying side-remark:--- On the other hand, I see no worries, even for Bricker 
and other advocates of a principle of recombination, lurking in the fact that laws 
typically require the values of quantities, including scalar intrinsic  quantities 
(such as temperature in my example), to be continuous, or even differentiable a 
specified number of times. For the principle says only that a world given by 
recombination is logically possible---not that it obeys the laws of either of the 
worlds of the ``recombined ingredients''.} Besides, I would say that  only someone 
in the grip of {\em pointillisme}---an explicit advocate like Lewis, or someone 
feeling its lure---would be uneasy at having fundamental (if you like, in Lewisian 
terms: perfectly natural) quantities that are extrinsic to a point. And so much 
the worse for {\em pointillisme}!

But to return to Bricker: he believes that both horns of the dilemma are 
unacceptable---and so his own preference is ...

\subsection{A heterodox but intrinsic metric}\label{sssec;heterodox}
Bricker thinks we should retain his original metaphysical framework, with its 
claim that all perfectly natural properties are intrinsic; and we should  escape 
the contradiction at the end of Section \ref{sssec;metrextr}, by giving up the 
idea that the metric is a perfectly natural property {\em of a point}. That is, he 
proposes:\\
\indent\indent (Heterodox):  {\em We should take the metric tensor to represent an 
intrinsic property of an infinitesimal neighbourhood of a point, taking such 
neighbourhoods as genuine mathematical objects.} \\
More precisely, we should hold that the metric, an extrinsic and {\em not} 
perfectly natural property of a point, is ``grounded'' in another intrinsic, 
perfectly natural property of a neighbourhood (Bricker's scare-quotes).

Since Bricker presents this preferred solution briefly, and I shall object to it, 
it is both clearest and fairest to quote him at length.  He writes
\begin{quote} 
To illustrate the sort of grounding I have in mind, consider mass density. If one 
assumes that each neighbourhood of a point has some determinate (finite) mass and 
volume, then the {\em mass density} at a point can be characterized as the limit 
of the ratio of mass to volume, as volume shrinks to zero. So characterized, mass 
density is an extrinsic property of points. But it is customary in physics, when 
considering a continuous matter field, to instead take mass density to be a 
primitive scalar field: a function that assigns to each point a real number 
representing (given appropriate units) the {\em intrinsic mass density} at the 
point. Given intrinsic mass density, and an assumption about its smooth 
distribution, mass can be defined by integration. Extrinsic mass density thus 
supervenes upon intrinsic mass density. And, thanks to a fundamental theorem of 
integral calculus, the values of extrinsic and intrinsic mass density coincide 
\ldots The suggestion, then, is to say something analogous about the local metric: 
the extrinsic local metric supervenes on an intrinsic local metric (plus manifold 
structure).(1993, p. 290-1.) 
\end{quote}
But, he then says, there is a problem.
\begin{quote} How can a tensor be intrinsic to a {\em point}? Points are spatially 
simple. Tensors, being operators on vector spaces, are spatially complex. It is 
repugnant to the nature of a point to suppose that a local metric, which is a 
tensor, could be intrinsic to a point \ldots [the intrinsic local metric] had 
better be intrinsic not to a point, but to something spatially complex. (1993, p. 
291.; with a footnote endorsing Robinson's argument which I reported in Section 
\ref{sssec;focusBric}, that vectorial properties must be extrinsic to a point.)
\end{quote}
He immediately goes on
\begin{quote} No sooner said than done. If we are willing to postulate perfectly 
natural properties  on theoretical grounds, we should be willing to posit 
appropriate entities to instantiate those properties: in this case, entities that 
are spatially complex. I propose that we reify talk of the ``infinitesimal 
neighbourhood'' of a point. The tangent space at a point is now conceived as the 
infinitesimal neighbourhood of the point ``blown large'' \ldots it no longer 
depends for its existence on the manifold structure. Tensor quantities are 
intrinsic \ldots to the infinitesimal neighbourhoods of points. \ldots space (or 
spacetime) has a ``non-standard'' structure. There are ``standard'' points, and 
there are ``non-standard'' points that lie an infinitesimal distance from standard 
points. The points along a path in space are ordered like the non-standard 
continuum of Abraham Robinson's non-standard analysis (ibid.)
\end{quote}

\subsection{Anti-{\em pointilliste} reply}\label{sssec;apreply}
My reply is clear from what I said in Section \ref{344B;vsintrdupls}. Namely, I 
think Bricker's principle of recombination is a poor reason for proposing 
non-standard analysis. Though of course non-standard analysis is impressive and 
fascinating, the fact that vectorial and tensorial properties are extrinsic to a 
point gives no good reason to adopt non-standard analysis as a metaphysical 
foundation for differential geometry: only {\em pointillisme} makes one think so.

The errors of {\em pointillisme} also show up in what Bricker says about his 
motivating example, mass density; in particular, his saying `it is customary in 
physics, when considering a continuous matter field, to instead take mass density 
to be a primitive scalar field'.

 I reply that this is a mistake. That is: the classical mechanics of continua 
(whether fluids or deformable solids) conceives mass density  exclusively as a 
limit of a ratio of mass to volume, and so as extrinsic---in just the way Bricker 
says at the start of the first quotation. And it is {\em right} to do so. For use 
of a primitive mass density scalar field leads to conceptual conundrums. (Agreed, 
under suitable conditions of smoothness, such a field meshes as regards the 
mathematics with the usual definition as a limit---as Bricker mentions.) 

As a very simple example of such a conundrum, imagine that the unit square 
$[0,1]^2 \subset \mathR^2$ is a sheet of continuous material, with a uniform mass 
density $\rho(x,y) = 1$ (so that the total mass is also 1). Now suppose the 
material is expanded to four times its original area, by a uniform stretch, so as 
to cover the set $[0,2]^2$. That is, there is a stretching function $f$:
\be
f: (x,y) \in [0; 1]^2   \mapsto  (2x, 2y) \in [0; 2]^2
\ee
The conservation of mass requires that after the expansion $\rho(x,y) = 0.25$ for 
all $(x,y)$. But if as Bricker suggests, the mass density $\rho$ is primitive, it 
is natural to ask: how does the point-sized bit of matter at a point $(x,y)$ 
``know'' how to decrease its value of $\rho$ between the initial and final times, 
say $t_0$ and $t_1$:  $\rho(x,y;t_1) = \frac{1}{4}\rho(x,y;t_0)$? After all, each 
point is mapped by $f$ to just one point, not to four points! 

On the other hand, there is no such conundrum (in this example, or countless 
others) if we first state the conservation of mass in terms of extended regions, 
and then treat mass density as a derived concept. 

In fact, we have here  come full circle: we have returned to Section 
\ref{sssec;line}'s  paradox of the length of a line, which launched our discussion 
of whether {\em pointillisme} can accommodate the structure of space or spacetime. 
For rigorous presentations of continuum  mechanics (e.g. Truesdell 1991, pp. 
16-19, 92-94) treat mass and mass density in exactly this way. That is: they 
postulate a mass measure that assigns values of mass to (an appropriate subset of 
all) spatial regions. Mass density is then introduced as a derived concept 
(essentially the limit of the ratio of mass to volume, as mentioned above), 
subject to certain conditions that ensure that its integral yields back the 
original mass-measure for  regions.  The full details of this treatment require 
modern measure theory: (for example, the conditions for the density's integral to 
equal the original measure are given essentially by the Radon-Nikodym theorem; for 
details, cf. e.g. Kingman and Taylor (1977, Theorem 6.7)). 

But I do not need to rehearse these details: here it is enough to give a 
non-rigorous statement of how this treatment, applied to the unit square example, 
avoids the conundrum of how $\rho$ can ``know'' how to decrease.\footnote{I also 
admit that (as mentioned in Section \ref{Intr} and \ref{sssec;line}) measure 
theory has some well-nigh paradoxical results of its own. But neither swallowing 
those results, nor avoiding them by revising measure theory, gives any support to 
{\em pointillisme}.}  

In this example, we postulate that all regions $R$ of a suitable kind $\cal K$ are 
assigned a mass $m(R,t)$ at a time $t$, which is conserved under the stretching in 
the sense that 
\be
\forall R \in {\cal K}, \;\;  m(R, t_0) = m(f(R), t_1).
\label{eq;consmass}
\ee
We also postulate that each region $R$ is assigned an area $a(R)$; and that the 
kind $\cal K$ is rich enough in the sense that for each point $(x,y)$, there is a 
sequence of regions $\{R_n\}$ which all contain $(x,y)$ but whose areas descend to 
0---which we write as $R_n \rightarrow (x,y)$. Then we  define the mass density at 
$(x,y)$ as the corresponding  limit of the ratio of mass to area: we assume here 
that this limit exists, for all $(x,y)$. That is:
\be 
\rho(x,y; t) := {\rm lim}_{R_n \rightarrow (x,y)} \frac{m(R, t)}{a(R)}.
\ee
The conservation of mass, represented fundamentally by eq. \ref{eq;consmass}, can 
then be re-expressed in terms of the integral of the density
\be
	\int_R \rho(x,y;t_0) \;  dxdy   = \int_{f(R)} \rho(x, y; t_1)\; dxdy
\ee
And from this, it follows that $\rho$ must decrease uniformly by a factor of 4: 
i.e. $\rho(x,y;t_0) = 4\rho(x,y;t_1)$. That is ``how $\rho$ knows'' how to 
decrease!\footnote{Of course, conundrums like this about the unit square can be 
formulated not only about mass and mass density, but about arbitrary measures and 
their densities. And the solution provided by modern measure theory is the same: 
take the measure, with its assignment to extended regions, as primary, and take 
the density  as a derived concept, viz. as a limit of the ratio of the measure to 
a volume or area measure.}

{\em Acknowledgements}:---  I am grateful to audiences at Florence, Kirchberg, 
Leeds, London, Oxford, and Princeton; and to A. Elga, G. Belot, P. Forrest, J. 
Hawthorne, S. Leuenberger,  L. Lusanna,  M. Pauri, H. Price, J. Uffink, and B. van 
Fraassen for conversations and comments.

\section{References}
Arntzenius, F. (2004), `Gunk, topology and measure', available at: 
http://philsci-archive.pitt.edu/archive/00001792. \\ 
Arntzenius, F. and Hawthorne, J (2006), `Gunk and continuous variation', 
forthcoming.\\
Arthur, R. (2006) `Leibniz's syncategorematic infinitesimals, smooth infinitesimal 
analysis and Newton's proposition 6', forthcoming.\\
Batterman, R. (2003), `Falling cats, parallel parking and polarized light',  {\em 
Studies in the History and Philosophy of Modern Physics} {\bf 34B}, pp. 527-558. 
\\
Bell, J. (1998), {\em A Primer of Infinitesimal Analysis}, Cambridge University 
Press.\\
Belot, G. (1998), `Understanding electromagnetism', {\em British Journal for the 
Philosophy of Science} {\bf 49}, pp. 531-555.\\
Black, R. (2000), `Against Quidditism', {\em Australasian Journal of Philosophy} 
{\bf 78}, pp. 87-104.\\
Bricker, P. (1993), `The fabric of space: intrinsic vs. extrinsic distance 
relations', in ed.s P.French et al., {\em Midwest Studies in Philosophy} {\bf 18}, 
University of Minnesota Press, pp. 271-294.\\
Butterfield, J. (2004), `On the Persistence of Homogeneous Matter', available at: 
physics/0406021: and at http://philsci-archive.pitt.edu/archive/00002381/ \\
Butterfield, J. (2004a), `The Rotating Discs Argument Defeated', forthcoming in 
{\em British Journal for the Philosophy of Science}; available at:\\ 
http://philsci-archive.pitt.edu/archive/00002382/ \\
Butterfield, J. (2005)  `On the Persistence of Particles', in {\em Foundations of  
Physics} {\bf 35}, pp. 233-269, available at: physics/0401112; and 
  http://philsci-archive.pitt.edu/archive/00001586/.\\
Butterfield, J. (2006), `Against {\em Pointillisme} about mechanics', forthcoming 
in {\em British Journal for the Philosophy of Science}; available at 
http://philsci-archive.pitt.edu. \\
Butterfield, J. (2006a), `Against {\em Pointillisme}: a call to arms', in 
preparation. \\
Butterfield, J. and Isham, C. (1999), `On the Emergence of Time in Quantum 
Gravity', in ed. J Butterfield, {\em The Arguments of Time}, British Academy and 
Oxford University Press, pp. 111-168; available at: gr-qc/9901024.\\
Butterfield, J. and Isham, C. (2001), `Spacetime and the Philosophical Challenge 
of Quantum Gravity', in ed.s  C. Callender and N. Huggett, {\em Physics meets 
Philosophy at the Planck Scale}, Cambridge University Press pp. 33-89; available 
at: gr-qc/9903072. \\
Caratheodory, C. (1963) {\em Algebraic Theory of Measure and Integration}, trans. 
F. Linton, New York: Chelsea Publishing Company.\\
Davies, E. (2003)  `Quantum mechanics does not require the continuity of space', 
{\em Studies in the History and Philosophy of Modern Physics} {\bf 34B}, pp. 
319-328.\\
Earman (1987), `Locality, non-locality and action-at-a-distance: a skeptical 
review of some philosophical dogmas', in {\em Kelvin's Baltimore Lectures and 
Modern Theoretical Physics},  eds. R. Kargon and P. Achinstein, Cambridge Mass: 
MIT Press.\\
Earman and Roberts (2006), `Contact with the Nomic: a challenge for deniers of 
Humean supervenience about laws of nature', {\em Philosophy and Phenomenological 
Research} forthcoming. \\
Forrest, P. (2002), `Non-classical mereology and its application to sets', {\em 
The Notre Dame Journal of Formal Logic} {\bf 43}, pp. 79-94.\\
Forrest, P. (2004), `Grit or gunk: implications of the Banach-Tarski paradox', 
{\em The Monist} {\bf 87}, pp. 351-384. \\
Hawthorne, J. (2006), `Quantity in Lewisian Metaphysics', forthcoming in his {\em 
Metaphysical Essays}, Oxford University Press.  \\
Hoefer, C. (2000), `Energy Conservation of in GTR', {\em Studies in the History 
and Philosophy of Modern Physics}, {\bf 31B} pp. 187-200. \\
Kingman, J. and Taylor, S. (1977), {\em Introducton to Measure and Probability}, 
Cambridge University Press.\\
Kragh, H. and Carazza, B. (1994), `From time atoms to spacetime quantization: the 
idea of discreet time 1925-1926', {\em Studies in the History and Philosophy of 
Modern Physics} {\bf 25}, pp. 437-462.\\ 
Langton, R. and Lewis, D. (1998), `Defining `intrinsic'', {\em Philosophy and 
Phenomenological Research} {\bf 58}, pp. 333-345; reprinted in Lewis (1999a), page 
reference to reprint.\\
Leibniz, G. (2001), {\em The Labyrinth of the Continuum: writings on the continuum 
problem 1672-1686}, ed. sel. and transl. R. Arthur, New Haven: Yale University 
Press.\\
Lewis, D. (1983), `Extrinsic properties', {\em  Philosophical Studies} {\bf 44}, 
pp. 197-200; reprinted in Lewis (1999a); page references to reprint.\\
Lewis, D. (1983a), `New Work for a Theory of Universals', {\em Australasian 
Journal of Philosophy} {\bf 61}, pp. 343-77;  reprinted in Lewis (1999a), page 
reference to reprint.\\
Lewis, D. (1986), {\em Philosophical Papers, volume II}, New York: Oxford 
University Press.\\
Lewis, D. (1986a), {\em On the Plurality of Worlds}, Oxford: Blackwell.\\
Lewis, D. (1994), `Humean Supervenience Debugged', {\em Mind} {\bf 103}, p 
473-490; reprinted in Lewis (1999a), pp. 224-247; page reference to reprint.\\
Lewis, D. (1999), `Zimmerman and the Spinning sphere', {\em Australasian Journal 
of Philosophy} {\bf 77}, pp. 209-212.\\
Lewis, D. (1999a), {\em Papers in Metaphysics and Epistemology}, Cambridge:  
University Press.\\
Lewis, D. (2001), `Redefining `intrinsic'', {\em Philosophy and Phenomenological 
Research} {\bf 63}, pp. 381-398.\\ 
McLarty, C. (1988), `Defining sets as sets of points of spaces', {\em Journal of 
Philosophical Logic} {\bf 17}, pp. 75-90.\\
Mancosu, P. (1996), {\em Philosophy of Mathematics and Mathematical Practice in 
the Seventeenth Century}, Oxford: University Press.\\
Menger, K. (1978), `Topology without points', in {\em Selected papers in Logic and 
Foundations, Didactics, Economics}, Dordrecht: Reidel, pp. 80-107.\\
Monk, N. (1997), `Conceptions of spacetime: problems and possible solutions, {\em 
Studies in the History and Philosophy of Physics} {\bf 28}, pp. 1-34. \\
Nelson, E. (1987), {\em Radically Elementary Probability Theory}, Princeton 
University Press.\\
Oppy, G. (2000), `{\em Humean} supervenience?', {\em Philosophical Studies} {\bf 
101} pp. 77-105.\\
Robinson, A. (1996), {\em Non-standard Analysis}, Princeton University Press.\\
Robinson, D. (1989), `Matter, Motion and Humean supervenience', {\em Australasian 
Journal of Philosophy} {\bf 67},  pp. 394-409.\\
Roeper, P. (1997), `Region-based topology' {\em Journal of Philosophical Logic} 
{\bf 26}, pp. 251-309.\\
Sider, T.  (2001), {\em Four-Dimensionalism}, Oxford University Press.\\
Skyrms, B. (1993), `Logical atoms and combinatorial possibility', {\em Journal of 
Philosophy} {\bf 90}, pp. 219-232.\\
Taylor, B. (1993), `On natural properties in metaphysics', {\em Mind} {\bf 102}, 
pp. 81-100.\\
Truesdell, C. (1991), {\em A First Course in Rational Continuum Mechanics}, volume 
1; second edition; Academic Press. \\ 
Vallentyne, P. (1997), `Intrinsic properties defined', {\em Philosophical Studies} 
{\bf 88}, pp. 209-219.\\
Wagon, S. (1985), {\em The Banach-Tarski Paradox}, Cambridge: University Press.\\
Weatherson, B. (2002), `Intrinsic vs. extrinsic properties', {\em Stanford 
Encyclopedia of Philosophy}, http://plato.stanford.edu/intrinsic-extrinsic.\\
Wilson, M (1998), `Classical mechanics'; in {\em The Routledge Encyclopedia of 
Philosophy}.\\
D. Zimmerman (1998), `Temporal parts and supervenient causation: the 
incompatibility of two Humean doctrines', {\em Australasian Journal of Philosophy} 
{\bf 76}, pp. 265-288.

\end{document}